\documentclass[useAMS,usenatbib]{mn2e}
\usepackage{graphicx}

\title[HD\,3651\,B: the first brown dwarf companion of an exoplanet host star imaged directly]
{HD\,3651\,B: the first directly imaged brown dwarf companion of an exoplanet host star}
\author[Mugrauer et al.]{M. Mugrauer$^{1}$\thanks{E-mail: markus@astro.uni-jena.de}, A. Seifahrt$^{1, 2}$, R. Neuh\"auser$^{1}$, T. Mazeh$^{3}$
\thanks{Based on observations obtained on La Silla in ESO programs 073.C-0103(A) and 077.C-0572(A),
and on Mauna Kea in UKIRT program U/02A/16.}\\
$^{1}$Astrophysikalisches Institut, Universit\"at Jena, Schillerg\"a{\ss}chen 2-3, 07745 Jena, Germany\\
$^{2}$European Southern Observatory, Karl-Schwarzschild-Str. 2, 85748 Garching, Germany\\
$^{3}$Tel Aviv University, Tel Aviv 69978, Israel \\}

\begin{document}

\date{Accepted 2006 August 23 , Received 2006 August 23 ; in original form 2006
July 28}

\pagerange{\pageref{firstpage}--\pageref{lastpage}} \pubyear{2006}

\maketitle

\label{firstpage}

\begin{abstract}

In the course of our ongoing multiplicity study of exoplanet host stars we detected a faint
companion located at $\sim$~43\,arcsec (480\,AU physical projected separation) north-west of its
primary --- the exoplanet host star HD\,3651 at 11\,pc. The companion, HD\,3651\,B, clearly shares
the proper motion of the exoplanet host star in our four images, obtained with ESO/NTT and UKIRT,
spanning three years in epoch difference. The magnitude of the companion is $H=16.75\pm0.16$\,mag,
the faintest co-moving companion of an exoplanet host star imaged directly. HD\,3651\,B is not
detected in the POSS-II B-, R- and I-band images, indicating that this object is fainter than
$\sim$~20\,mag in the B- and R-band and fainter than $\sim$~19\,mag in the I-band. With the
Hipparcos distance of HD\,3651 of 11\,pc, the absolute magnitude of HD\,3651\,B is about 16.5\,mag
in the H band. Our H-band photometry and the Baraffe et al. (2003) evolutionary models yield a mass
of HD\,3651\,B to be 20 to 60\,$M_{Jup}$ for assumed ages between 1 and 10\,Gyr. The effective
temperature ranges between 800 and 900\,K, consistent with a spectral type of T7 to T8. We conclude
that HD\,3651\,B is a brown-dwarf companion, the first of its kind directly imaged as a companion
of an exoplanet host star, and one of the faintest T dwarfs found in the solar vicinity (within
11\,pc).
\end{abstract}


\section{Introduction}

During the last decade high precision radial-velocity studies revealed almost 200 exoplanets around
mainly solar-like stars located in the solar neighborhood. The stellar multiplicity of these planet
host stars was already investigated by several groups using imaging data from visible and infrared
all sky surveys like POSS or 2MASS, see e.g. Bakos et al. (2006) or Raghavan et al (2006). With
seeing limited near infrared imaging (see e.g. Mugrauer et al. 2005 and 2006) as well as high
contrast diffraction limited AO observations (e.g. Els et al. 2001, Patience et al. 2002, Luhman \&
Jayawardhana 2002, and Chauvin et al. 2006) further companions of exoplanet host stars have been
detected during the last few years. Up to now, more than 30 companions of exoplanet host stars are
known, suggesting that the multiplicity of these stars is at least 20\,\%.

Faint companions of exoplanet host stars were detected close to HD\,114762 (Patience et al. 2002)
and Gl\,86 (Els et al. 2001) with angular separations smaller than 3\,arcsec, which corresponds to
physical projected separations of 130 and $\sim$~20\,AU, respectively. In addition to these faint
close objects, Wilson et al. (2001) reported a faint, wide companion to HD\,89744, detected in the
2MASS point source catalogue (Skrutskie et al. 2006). HD\,89744\,B is located 63\,arcsec ($\sim$
2500\,AU physical projected separation) north-east of HD\,89744\,A, and shares the proper motion of
this exoplanet host star (Mugrauer et al. 2004). While the faint companions HD\,114762\,B
($M_{H}\sim10.4$\,mag) and HD\,89744\,B ($M_{H}\sim11.1$\,mag) are both very low-mass objects with
masses of about 0.080\,$M_{\sun}$, Gl\,86\,B ($M_{H}\sim14.2$\,mag) turned out to be a white dwarf
(see Mugrauer \& Neuh\"auser 2005 for further details).

So far, substellar companions with masses significantly below the stellar/substellar mass border of
0.075\,$M_{\sun}$ have not been directly detected as companions to rad-vel planet host stars,
neither with AO nor seeing limited imaging. In contrast, radial-velocity studies have already
revealed a number of brown-dwarf companions close to the exoplanet host stars e.g. HD\,38529
(Fischer et al. 2003a), HD\,168443 (Marcy et al. 2001), and HD\,202206 (Correia et al. 2005), with
minimum masses of $msin(i)=12.7$, 17.2 and 17.4\,$M_{Jup}$, respectively. While the brown-dwarf
companions of HD\,38529 and HD\,168443 both revolve around the planet host star on orbits with
semi-major axes which are at least 10 times wider than those of the detected exoplanets, the
brown-dwarf companion of HD\,202206 moves around the planet host star inside the planetary orbit.
All of these systems are interesting examples which might point out that planet and brown-dwarf
formation processes can occur around the same star at comparable distances (see Correia et. al.
2005 for further discussion).

In this letter we present the first direct imaged brown-dwarf companion of an exoplanet host star
which we detected in the course of our multiplicity study. We show the results of our near infrared
observations in section\,2 and discuss the properties of this new very faint companion in
section\,3.

\section{Observations}

HD\,3651 is a high proper motion nearby K0 dwarf, located at the border of the northern
constellations Pegasus and Pisces. Its proper motion and parallax
($\mu_{\alpha}cos(\delta)=-461.09\pm0.75$\,mas/yr, $\mu_{\delta}=-370.90\pm0.61$\,mas/yr, and
$\pi=90.03\pm0.72$\,mas) are determined by Hipparcos (Perryman et al.  1997), yielding a distance
of $\sim$~11\,pc. According to Fischer et al. (2003b) HD\,3651 is chromospheric inactive and
photometrically stable, as expected for a middle-aged early K dwarf. The stellar radial velocity
shows a periodical modulation with a period of 62.23\,day, indicating the star is orbited by a
sub-Saturn mass planet ($msin(i)=0.20$\,$M_{Jup}$) with an eccentric ($e=0.63$) orbit and a
semi-major axis of $a=0.284$\,AU. According to Santos et al. (2004) HD\,3651 exhibits an effective
temperature of $5173\pm35$\,K and a surface gravity of $log(g)=4.37\pm0.12$\,cm/s$^{2}$, as it is
expected for an early K dwarf. Its metallicity is slightly enhanced compared to the sun
($[Fe/H]=0.12\pm0.04$), typical for a planet host star. Santos et al. (2004) also derive the mass
of HD\,3651 to be 0.76\,$M_{\sun}$, consistent with the mass estimate of 0.79\,$M_{\sun}$ from
Fischer et al. (2003b).

The age of this exoplanet host star has been estimated by several groups. Valenti et al. (2005)
report an age between 3 and 12.5\,Gyr, derived with isochron fitting. This range is consistent with
the chromospheric age estimates of 5.9\,Gyr from Wright et al. (2004) and 2.1\,Gyr from Rocha-Pinto
et al. (2004).

HD\,3651 is a target of our imaging search campaign for visual companions of northern exoplanet
host stars which was carried out with the 3.8\,m United Kingdom Infrared Telescope (UKIRT), located
at Mauna Kea (Hawaii). We observed HD\,3651 in the H-Band with the infrared camera UKIRT Fast Track
Imager (UFTI) a 1024\,$\times$\,1024 HgCdTe infrared detector with a pixel scale of $\sim$~91\,mas
per pixel and 93\,arcsec\,$\times$\,93\,arcsec field of view. The first epoch imaging is carried
out in June 2003. In order to reduce strong saturation effects of the bright exoplanet host star
the shortest available integration time (4\,s) was used and six of these 4\,s integrations were
averaged to one image. The standard jitter/dither technique was applied to minimize the sky
background.

In total 22 jitter positions were chosen, resulting in a total integration time of 8.8\,min. The
ESO package \textsl{ECLIPSE} (Devillard 2001) was used for background estimation and subtraction as
well as the flat fielding of all images which where finally combined by shift+add.

In addition to HD\,3651 we observed many other exoplanet host stars in that run. With all these
images we can derive the astrometric calibration of the UFTI detector using the 2MASS point source
catalogue (Skrutskie et al. 2006), which contains accurate positions of objects brighter than
15.2\,mag in H ($S/N$\,$>$\,5). The derived pixel scale and detector orientation of all observing
runs presented here are summarized in Tab.\,1. The achieved detection limit ($S/N>10$) of the UFTI
observations is H$\sim$18.4\,mag which is reached in the background noise limited region at angular
separations larger than 15\,arcsec around the bright exoplanet host star.

\begin{figure*}\centering\includegraphics[width=17.5cm]{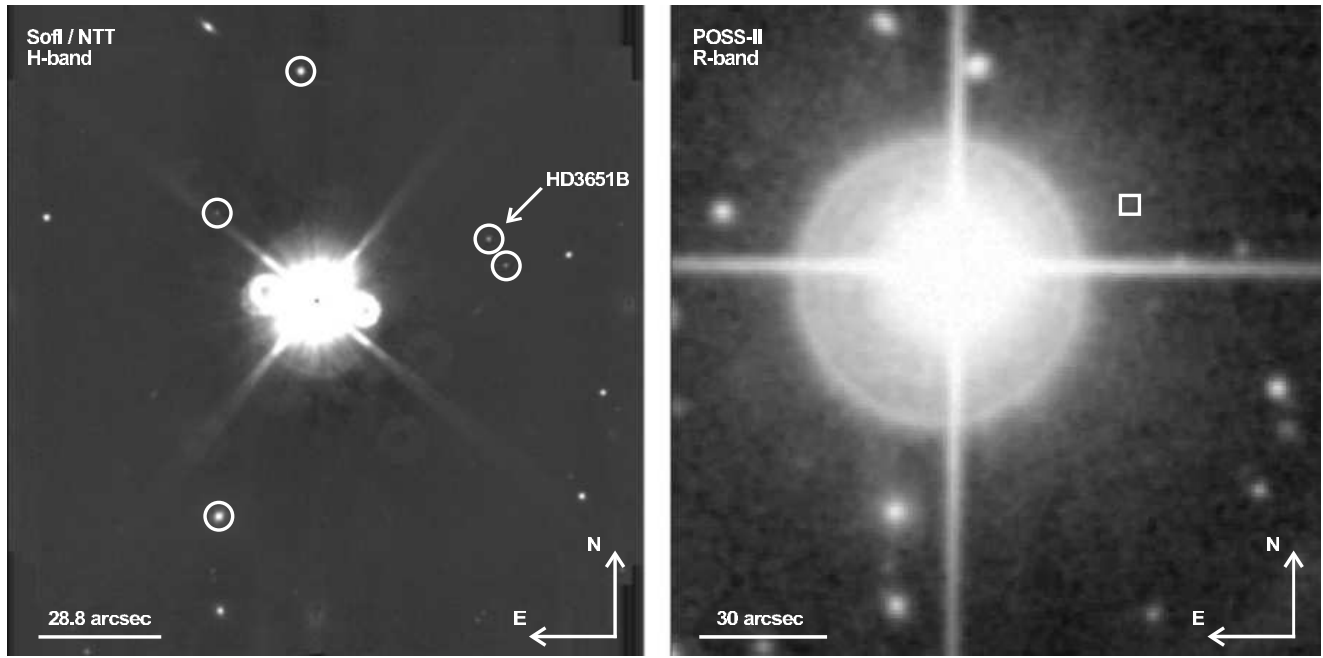}
\caption{Left pattern: The SofI small field image of the planet host star HD\,3651, taken in June
2006 in the H-band. All companion candidates which are also detected in our UFTI H-band images are
marked with white circles. The co-moving companion HD\,3651\,B is indicated with a black arrow.
Right pattern: The exoplanet host star HD\,3651 detected on the R-band POSS-II photographic plate
from observing epoch August 1987. Faint objects are detected around HD\,3651 with magnitudes down
to $\sim$~21\,mag. The expected position of HD\,3651\,B is illustrated with a white box but the
companion is not detected in the POSS image. The motion of the exoplanet host star relative the
background sources is already visible by comparing the POSS image with our SofI image.}
\label{hd3651pic}
\end{figure*}

\begin{table*}
\begin{center}
\caption{The pixel scale and the detector orientation with their uncertainties as well as the
average seeing for all UFTI and SofI observing runs. The detector is tilted by the given angle from
north to west. Furthermore, the separations and position angles of HD\,3651\,B relative to its
primary --- the exoplanet host star HD\,3651 and the H-band photometry of the companion are listed,
as measured in all UFTI and SofI observing epochs.}
\begin{tabular}{lcccccc}
\hline
Epoche & Pixel Scale & Detector Orientation & seeing & separation & position angle & H magnitude\\
Camera \& Date& $[$arcsec/Pixel$]$ & $[$$^{\circ}$$]$ & $[$arcsec$]$ & $[$arcsec$]$ & $[$$^{\circ}$$]$ & $[$mag$]$\\
\hline
UFTI 06/03 & 0.09109$\pm$0.00035 & 0.747$\pm$0.096 & 0.66 & 42.92$\pm$0.17 & 289.92$\pm$0.24 & 16.58$\pm$0.07\\
UFTI 10/03 & 0.09104$\pm$0.00030 & 0.711$\pm$0.083 & 0.69 & 43.15$\pm$0.14 & 289.99$\pm$0.21 & 16.83$\pm$0.07\\
SofI\,\,\,\, 07/04 & 0.14356$\pm$0.00011 & 90.047$\pm$0.024 & 1.25 & 42.89$\pm$0.08 & 289.92$\pm$0.10 & 16.93$\pm$0.10\\
SofI\,\,\,\, 06/06 & 0.14348$\pm$0.00016 & 90.017$\pm$0.049 & 0.73 & 43.07$\pm$0.08 & 289.97$\pm$0.11 & 16.65$\pm$0.07\\
\hline
\end{tabular}
\end{center}
\label{data}
\end{table*}

Five faint objects around HD\,3651 are detected in our UFTI image, all of which might be real
companions of the exoplanet host star. With a follow-up 2nd epoch observation we tested the
companionship of the detected companion-candidates. Real companions of the exoplanet host star
would share the proper motion of their primary, as their orbital motion is much smaller than the
proper motion of the exoplanet host star. Such co-moving companions can therefore be easily
distinguished from non- or slowly moving background stars by comparing two images taken with
sufficient long time difference.

The 2nd epoch UFTI observation of HD\,3651 was obtained in October 2003. We averaged again six 4\,s
integrations per jitter-position but chose this time a 24 position jitter pattern, yielding a total
integration time of 9.6\,min.

Although the time difference between the two UFTI observations is only 4 months, the astrometric
companion search is already feasible because of the high proper motion of HD\,3651, which amounts
to 0.6\,arcsec/yr. All candidates except one significantly change their separations and position
angles relative to the exoplanet host star. Only one candidate does not show a significant
variation in its position angle and separation. Its relative astrometry is summarized in Tab.\,1,
and is illustrated in Fig.\,2. This object might be a real companion of the exoplanet host star,
and therefore is being denoted as HD\,3651\,B henceforth.

\begin{figure}\resizebox{\hsize}{!}{\includegraphics{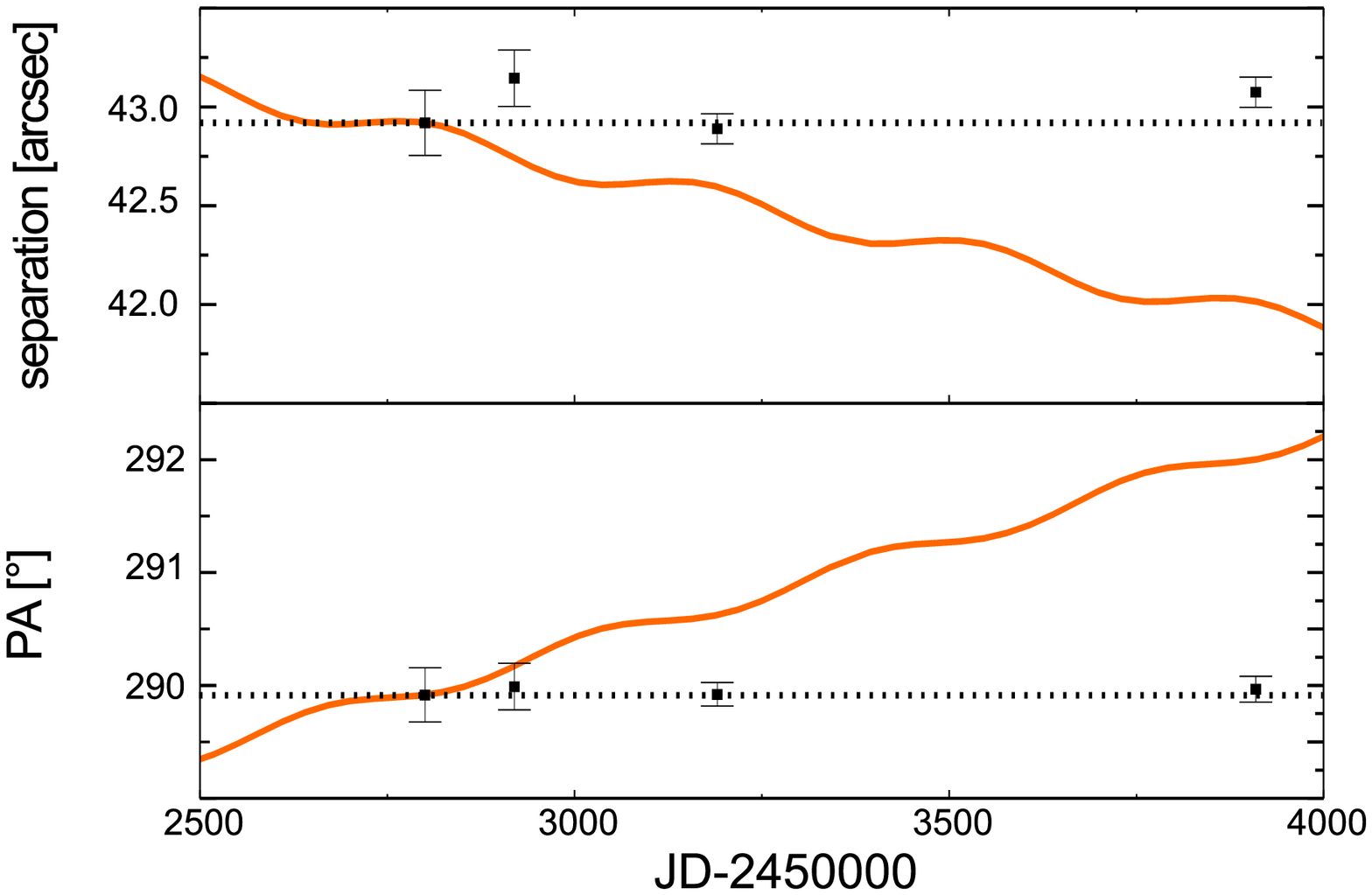}}
\caption{The separations and position angles of HD\,3651\,B relative to HD\,3651 for all UFTI and
SofI observing epochs. The solid line indicates the expected variation of separation and position
angle in case HD\,3651\,B is a non-moving background star, calculated with the Hipparcos proper and
parallactic motion of the exoplanet host star.} \label{seppa}
\end{figure}

Due to the proximity of HD\,3651, which is located only at a distance of 11\,pc, the UFTI field of
view allows only the detection of companions with physical projected separations up to
$\sim$~500\,AU from the exoplanet host star. In order to reach wider companions we include HD\,3651
in our companion search program of southern exoplanet host stars and nearby stars, both being
carried out at La Silla observatory with the ESO 3.58\,m New Technology Telescope (NTT) and its
infrared camera SofI, a 1024\,$\times$\,1024 HgCdTe-detector. All observation were obtained again
in the H-band using the SofI small field objective with its pixel scale of 144\,mas per pixel,
yielding a field of view of 147\,$\times$\,147\,arcsec. This allows the detection of wide
companions around HD\,3651 with physical projected separations up to $\sim$~800\,AU. Our 1st epoch
SofI observation of HD\,3651 was carried out in July 2004. The 2nd epoch follow-up imaging was
recently taken in June 2006. In both observing epochs we used the jitter technique and obtain 10
images, each the average of 50 1.2\,s integrations, i.e. 10\,min of total integration time. Our 2nd
epoch SofI image is shown in the left pattern of Fig.\,1.

Both SofI images are calibrated with the 2MASS point source catalogue (see Tab.\,1). HD\,3651\,B
and the 4 objects already detected in the UFTI images as well as several more wider companion
candidates are found in the SofI images. However, all of these additional companion candidates
exhibit only negligibly small proper motions, determined by the comparison of both SofI images. In
contrast, HD\,3651\,B clearly shares the proper motion of HD\,3651\,A and its separation and
position angle does not change during three years of epoch difference between the first UFTI
observation (epoch 06/03) and the 2nd epoch SofI imaging (epoch 06/06), hence this is a real
companion of the exoplanet host star (see Fig.\,2).

The reached detection limit of our SofI observations for a range of separations to the planet host
star is shown in Fig.\,3. We are sensitive to companions with apparent H-Band magnitudes of
17.5\,mag in the background noise limited region, i.e. at angular separations larger 16\,arcsec
($\sim$ 180\,AU) around the bright exoplanet host star. According to the Baraffe et al. (2003)
models and the magnitude-mass relation therein, the achieved limiting magnitude allows the
detection of brown-dwarf companions with a mass m\,$\ge$\,37\,$M_{Jup}$ for an assumed age of the
primary of 5\,Gyr.  Additional stellar companions (m$\ge$75\,$M_{Jup}$) can be ruled out beyond
4.5\,arcsec ($\sim$~50\,AU) up to separations of $\sim$~760\,AU. All objects within 69\,arcsec
around the exoplanet host star are imaged in both SofI observing epochs but, expect HD\,3651\,B, no
further co-moving companions are detected.

\begin{figure}\resizebox{\hsize}{!}{\includegraphics{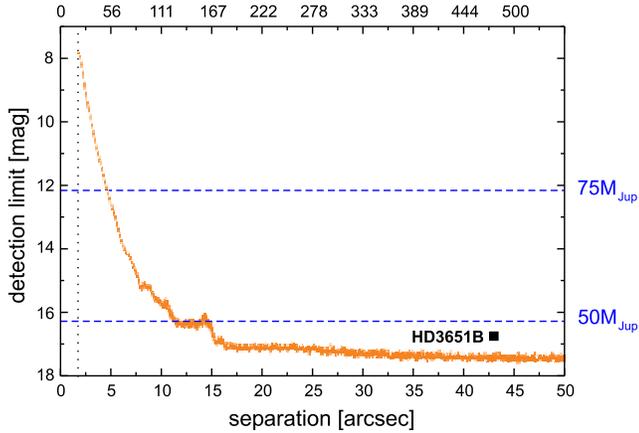}}
\caption{The detection limit ($S/N$\,=\,10) of the SofI H-band imaging of the exoplanet host star
HD\,3651 plotted for a range of separations given in arcsec at the bottom and as physical projected
separation in AU at the top. At $\sim$~1.8\,arcsec (20\,AU) saturation occurs (doted line). At a
system age of 5\,Gyr the achieved limiting magnitude allows the detection of substellar companions
with a mass m\,$\ge$\,37\,$M_{Jup}$ in the background limited region beyond 16\,arcsec
($\sim$~180\,AU). All stellar companions (m\,$\ge$\,75\,$M_{Jup}$) are detectable beyond the
distance illustrated by the upper dashed line ($\sim$~50\,AU). The faint co-moving companion
HD\,3651\,B is plotted as a black square.} \label{limit}
\end{figure}

We determine the apparent H-band magnitude of the co-moving companion HD\,3651\,B in all UFTI and
SofI images, all of which are listed in Tab.\,1. We measure the companion magnitude with aperture
photometry using the data reduction package ESO-MIDAS. The photometric zero points of the
observations are derived with sources detected in our UFTI and SofI images whose H-band magnitude
is listed in the 2MASS point source catalogue. The averaged apparent H-band magnitude of the
companion is $H=16.75\pm0.16$\,mag. The Hipparcos parallax $\pi=90.03\pm0.72$\,mas of the exoplanet
host star finally yields an absolute magnitude of $M_{H}=16.52\pm0.16$\,mag.

\section{On the nature of HD\,3651\,B}

Our H-band multi-epoch UFTI and SofI observations of the exoplanet host star HD\,3651 revealed the
faint co-moving companion HD\,3651\,B, separated from its primary by 43\,arcsec, corresponding to
physical projected separation of $\sim$~480\,AU. With an apparent H-band magnitude of
$H=16.75\pm0.16$\,mag, HD\,3651\,B is the faintest co-moving companion of an exoplanet host star
detected so far. Up to now the faintest known companion of an exoplanet host star is Gl\,86\,B,
whose absolute H-band magnitude was measured by Els et al. (2001) to be $M_{H}=14.2\pm0.2$\,mag.
With its absolute H-band magnitude of $M_{H}\sim16.5$\,mag, HD\,3651\,B is 2.3\,mag fainter than
Gl\,86\,B.

HD\,3651\,B is detected in our near infrared UFTI and SofI images but we do not find a visible
counterpart on the photographic B-, R-, and I-band plates of the 2nd Palomar All Sky Survey
(POSS-II). The POSS-II R-band plate is shown in the right pattern of Fig.\,1. According to Griffin
(2002) the detection limit of the POSS-II plates are 22.5\,mag in B-band, 20.8\,mag in R-band, and
19.5\,mag in I-band. Hence, HD\,3651\,B has to be fainter than $\sim$~20\,mag in R- and B-band and
fainter than $\sim$~19\,mag in the I-band to remain undetectable on all POSS plates.

HD\,3651\,B is a faint source in the near infrared H-band and it is not detectable in the B-, R-
and I-band. Due to its faintness in the optical spectral range compared to its near infrared H-band
photometry, HD\,3651\,B is not a cool white dwarf which is expected to be comparable bright in all
photometric bands (see Bergeron et al. 1995), i.e. should be detectable on the POSS plates. In
contrast, according to the evolutionary models of low-mass substellar objects from Baraffe et al.
(2003) all these photometric results are fully consistent with a cool brown dwarf located at the
distance of the exoplanet host star. With this models we can also derive the physical properties of
HD\,3651\,B from the measured H-band absolute magnitude assuming different system ages. In Tab.\,2
we have summarized the derived companion masses and effective temperatures as well as the expected
absolute R- and I-band magnitudes for assumed ages of 1, 5 and 10\,Gyr (B-band magnitudes not given
in theses models). The mass ranges from 22 up to 57\,$M_{Jup}$ for an age of 1 and 10\,Gyr,
respectively. The expected effective temperature of the brown dwarf ranges between 800 and 900\,K.
According to Golimowski et al. (2004) the derived temperature range is consistent with a spectral
type of T7. Furthermore, if we use the $M_{H}$ versus spectral type relation, derived by Vrba et
al. (2004), the absolute magnitude of HD\,3651\,B is consistent with a brown dwarf of spectral type
T7.9. Follow-up spectroscopy would be interesting to confirm the spectral type.

After $\epsilon$\,Ind\,B (McCaughrean et al. 2004) and SCR\,1845-6357\,B (Biller et al. 2006)
HD\,3651\,B is a further late T dwarf in the solar neighborhood. Its companionship to an exoplanet
host star is especially remarkable and puts news constraints to the formation theory of brown
dwarfs and extrasolar planets. This finding is a further evidence that both formation process can
occur around the same object.

\begin{table}
\begin{center}
\caption{The mass and effective temperature of HD\,3651\,B as well as its expected R- and I-band
magnitudes derived from our UFTI and SofI H-band photometry and the Baraffe et al. (2003)
evolutionary models.}
\begin{tabular}[h]{l|l|c|c|c}
\hline
age       & $[$Gyr$]$       & 1          & 5          & 10\\
\hline
mass      & $[$$M_{Jup}$$]$ & 22$\pm$1   & 46$\pm$2   & 57$\pm$2\\
$T_{eff}$ & $[$K$]$         & 798$\pm$25 & 855$\pm$26 & 885$\pm$26\\
$M_{R}$ & $[$mag$]$ & 22.80$\pm$0.10 & 22.86$\pm$0.08 & 22.90$\pm$0.08 \\
$M_{I}$ & $[$mag$]$ & 20.01$\pm$0.10 & 20.08$\pm$0.09 & 20.11$\pm$0.09 \\
\hline
\end{tabular}
\end{center}
\label{properties}
\end{table}

\label{lastpage}

\end{document}